\documentclass[12pt]{iopart}
\usepackage{iopams}

\expandafter\let\csname equation*\endcsname\relax

\expandafter\let\csname endequation*\endcsname\relax

\usepackage{amsmath,amsthm,amssymb}

\usepackage{breakurl}             

\usepackage{tikzit}
\usepackage{pgfplots}

\usepackage{hyperref}

\usepackage{xspace,enumerate,color,epsfig}
\usepackage{graphicx}
\graphicspath{{.}{./figures/}}
\usepackage{marvosym}
\usepackage{bm}

\usepackage{stmaryrd}
\usepackage{docmute}
\usepackage{keycommand}

\let\olddagger\dagger
\renewcommand{\dagger}{\ensuremath{\olddagger}\xspace}

\theoremstyle{definition}
\newtheorem{theorem}{Theorem}[section]

\newtheorem{definition}[theorem]{Definition}

\newtheorem{example*}[theorem]{Example*}
\newtheorem{examples*}[theorem]{Examples*}

\newtheorem{remark*}[theorem]{Remark*}

\usepackage[color]{changebar}

\usepackage{color}
\def\bR{\begin{color}{red}} 
\def\bB{\begin{color}{blue}}
\def\bM{\begin{color}{magenta}}
\def\bC{\begin{color}{cyan}}
\def\bW{\begin{color}{white}}
\def\bBl{\begin{color}{black}} 
\def\bG{\begin{color}{green}}
\def\bY{\begin{color}{yellow}}
\def\e{\end{color}\xspace}
\newcommand{\bit}{\begin{itemize}}
\newcommand{\eit}{\end{itemize}\par\noindent}
\newcommand{\ben}{\begin{enumerate}}
\newcommand{\een}{\end{enumerate}\par\noindent}
\newcommand{\beq}{\begin{equation}}
\newcommand{\eeq}{\end{equation}\par\noindent}
\newcommand{\beqa}{\begin{eqnarray*}}
\newcommand{\eeqa}{\end{eqnarray*}\par\noindent}
\newcommand{\beqn}{\begin{eqnarray}}
\newcommand{\eeqn}{\end{eqnarray}\par\noindent}

\tikzstyle{box}=[shape=rectangle, text height=1.5ex, text depth=0.25ex, yshift=0.5mm, fill=white, draw=black, minimum height=5mm, yshift=-0.5mm, minimum width=5mm, font={\small}]
\tikzstyle{black vertex}=[fill=black, draw=black, minimum size=2mm, shape=circle, inner sep=0mm]
\tikzstyle{gray vertex}=[fill=gray, draw=gray, minimum size=2mm, shape=circle, inner sep=0mm]
\tikzstyle{sr circle}=[fill=none, draw=red, shape=circle, tikzit fill=red, minimum size=2.5mm, very thick]
\tikzstyle{st circle}=[fill=none, draw=blue, shape=circle, tikzit fill=blue, minimum size=2.5mm, very thick]
\tikzstyle{si circle}=[fill=none, draw={rgb,255: red,0; green,145; blue,0}, shape=circle, tikzit fill={rgb,255: red,0; green,145; blue,0}, minimum size=2.5mm, very thick]
\tikzstyle{cnot ctrl}=[fill=black, draw=black, shape=circle, inner sep=0pt, minimum width=1.2mm, tikzit category=circuit]
\tikzstyle{cnot targ}=[fill=white, draw=white, shape=circle, tikzit category=circuit, label={center:$\oplus$}, inner sep=0pt, minimum width=2.1mm, tikzit fill={rgb,255: red,102; green,204; blue,255}, tikzit draw=black]
\tikzstyle{gate}=[fill=white, draw=black, shape=rectangle, minimum width=6mm, minimum height=6mm, font={\footnotesize}]
\tikzstyle{Z dot}=[inner sep=0mm, minimum size=2mm, shape=circle, draw=black, fill={rgb,255: red,160; green,255; blue,160}]
\tikzstyle{Z phase dot}=[minimum size=5mm, font={\footnotesize\boldmath}, shape=rectangle, rounded corners=2mm, inner sep=0.2mm, outer sep=-2mm, scale=0.8, tikzit shape=circle, draw=black, fill={rgb,255: red,160; green,255; blue,160}, tikzit draw=blue]
\tikzstyle{X dot}=[Z dot, shape=circle, draw=black, fill={rgb,255: red,220; green,0; blue,0}]
\tikzstyle{X phase dot}=[Z phase dot, tikzit shape=circle, tikzit draw=blue, fill={rgb,255: red,220; green,0; blue,0}, font={\footnotesize\color{white}\boldmath}]

\tikzstyle{gray edge}=[-, thick, draw={rgb,255: red,214; green,214; blue,214}]
\tikzstyle{black edge}=[-, thick]
\tikzstyle{dashed edge}=[-, dashed]
\tikzstyle{hadamard edge}=[-, color=blue, dashed, dash pattern=on 3pt off 1.5pt, thick]
\tikzstyle{brace edge}=[-, tikzit draw=blue, decorate, decoration={brace,amplitude=1mm,raise=-1mm}]

\input{steiner.tikzdefs}

\usepackage{hyperref}
\usepackage{longtable}
\usepackage{csvsimple}
\usepackage{subcaption}

\newcommand{\steinergauss}{\textsc{steiner-gauss}\xspace}
\newcommand{\steinerrec}{\textsc{steiner-gauss-rec}\xspace}
\newcommand{\steinerdown}{\textsc{steiner-down}\xspace}
\newcommand{\steinerup}{\textsc{steiner-up}\xspace}
\newcommand{\bitfield}{\ensuremath{\textrm{GF}(2)}\xspace}
\newcommand{\rowop}[2]{\tikz{%
\node at (0,0.5) {\scriptsize $R_{#1}\! := \!R_{#1} \!+\! R_{#2}$};
\draw [-latex] (-1.5,0) -- (1.5,0);
}\xspace}

\begin{document}

\title[CNOT Circuit Extraction for Topologically-constrained Quantum Memories]{CNOT Circuit Extraction for Topologically-constrained Quantum Memories}

\author{Aleks Kissinger}
\address{Radboud University Nijmegen}
\ead{aleks@cs.ru.nl}

\author{Arianne Meijer-van de Griend}
\address{Radboud University Nijmegen}
\ead{ariannemeijer@gmail.com}

\vspace{10pt}
\begin{indented}
\item[]May 2019
\end{indented}

\begin{abstract}
  Many physical implementations of quantum computers impose stringent memory constraints in which 2-qubit operations can only be performed between qubits which are nearest neighbours in a lattice or graph structure. Hence, before a computation can be run on such a device, it must be mapped onto the physical architecture. That is, logical qubits must be assigned physical locations in the quantum memory, and the circuit must be replaced by an equivalent one containing only operations between nearest neighbours. In this paper, we give a new technique for quantum circuit mapping (a.k.a. routing), based on Gaussian elimination constrained to certain optimal spanning trees called Steiner trees. We give a reference implementation of the technique for CNOT circuits and show that it significantly out-performs general-purpose routines on CNOT circuits. We then comment on how the technique can be extended straightforwardly to the synthesis of CNOT+Rz circuits and as a modification to a recently-proposed circuit simplification/extraction procedure for generic circuits based on the ZX-calculus.
\end{abstract}


\newpage

\section{Introduction}\label{sec:intro}

Quantum circuits give a \textit{de facto} standard for representing quantum computations at a low level. They consist of sequences of primitive operations, called quantum gates, applied to a register of quantum bits, or qubits. Increasingly, noisy intermediate-scale quantum (NISQ) computers with 10-80 qubits are becoming a reality. Popular physical realisations such as superconducting quantum circuits~\cite{reagor2018demonstration, ibmSystemOne, superconductingdelft} and ion traps~\cite{britton2012traps2d,iontrapoxford,iontrapcolorado,hensinger2006t} consist of qubits stored in the physical states of systems arranged in space, where two-qubit operations are typically only possible between pairs of adjacent systems. Hence, when it comes to actually running a quantum computation on these architectures, logical qubits must be mapped to physical memory locations, and the circuit must be modified to only consist of 2-qubit operations between adjacent qubits in the physical architecture. Na\"ively, this can be achieved by simply inserting swap gates to move a pair of qubits next to each other before each 2-qubit operation. However, this approach comes with an enormous overhead in terms of 2-qubit operations, each of which introduces a great deal more noise than a single qubit operation on most realistic architectures~\cite{iontrapoxford}. More sophisticated approaches incorporate techniques from computer aided design~\cite{zulehner2018efficient} and machine-learning~\cite{herbert2018using} in order to minimise the extra operations needed by making good choices of initial and intermediate memory locations for the qubits involved. Nevertheless, these are simply refinements of the basic `search and swap' approach. Most approaches only take the topological structure of the circuit into account (i.e.~which qubits are being acted upon) rather than semantic structure (i.e.~the unitary being implemented), and hence miss out on opportunities for more efficient circuit mapping.

We present a new approach to quantum circuit mapping based on constrained Gaussian elimination, and apply it in the simplest case of mapping CNOT circuits. The main idea is to modify a familiar technique for synthesising CNOT circuits using Guassian elimination in such that a way that primitive row operations (i.e. CNOT gates) are only allowed between certain rows corresponding to neighbouring qubits. Hence, non-local row operations must be propagated through intermediate rows. We then give a simple strategy for identifying and using appropriate intermediate rows based on certain minimal spanning trees called Steiner trees. Once we produce such a spanning tree, we use CNOTs to propagate row operations down toward the leaves then ultimately back up toward the root. The end result is CNOT circuit realising a given parity map involving only nearest-neighbour interactions. To measure the effectiveness of our approach, we produce many random CNOT circuits on 9, 16, and 20 qubits, containing between 3 and 256 CNOT gates, and map them onto 5 different graph topologies: $3\times 3$ and $4 \times 4$ square lattices, 16-qubit architectures of the IBM QX-5 and Rigetti Aspen devices, and the 20-qubit IBM Q20 Tokyo architecture. To compare the performance of our technique to general-purpose mapping techniques, we also map these CNOT circuits with the Rigetti QuilC compiler and \textsf{t$|$ket$\rangle$} by Cambridge Quantum Computing. We chose these tools because they scale well to our larger test circuits and give state of the art results on a large set of benchmark circuits published by IBM~\cite{CQCdepth}. Using these as a baseline, we find an average savings in 2-qubit gates of 48\% over QuilC and 36\% over \textsf{t$|$ket$\rangle$}.

Since CNOTs and single-qubit operations are universal for quantum computation, this already extends to a routine for mapping generic circuits: simply apply our routine to CNOT-only sub-circuits. However, only circuits containing long sequences of CNOT gates are likely to benefit from this na\"ive approach.
To address this issue, we will discuss in Section~\ref{sec:extensions} how the techniques we describe can be extended to synthesis of more general families of circuits. We note that our technique for CNOT circuits extends straightforwardly to circuits consisting of CNOT and Z-phase gates using the \textit{phase polynomial} representation of these circuits (see e.g.~\cite{amy2014polynomial}). We also suggest a technique for mapping universal circuits by modifying recently-proposed method by one of the authors~\cite{cliff-simp} based on the ZX-calculus~\cite{CD1}. The extraction method from~\cite{cliff-simp} has been implemented (without topological constraints) in the PyZX~\cite{pyzx} circuit optimiser, which has already been very successful in reducing T-count for general Clifford+T circuits~\cite{ZXtcount}.

The paper is structured as follows. In Section~\ref{sec:parity} we give a brief background on the theory of CNOT circuits and parity maps and provide the definitions of Steiner trees and descending Steiner trees, which we will use in our synthesis algorithm. The algorithm itself is described in Section~\ref{sec:algorithm}, for the simpler case where the CNOT connectivity graph contains a Hamiltonian path (as in all five of our benchmark architectures). In Section~\ref{sec:results}, we give the results of our CNOT mapping procedure in five different architectures and compare performance to the Quilc and \textsf{t$|$ket$\rangle$} compilers. We then describe two extensions in Section~\ref{sec:extensions}: to general graphs and to general circuits.



\paragraph{Related work.} Most existing circuit mapping techniques are based on searching for the optimal placement of swap gates and qubits. This can be described mathematically and solved with a general solver or temporal planner \cite{murali2019formal, venturelli2018compiling}. However, the search space for finding optimal swap gates is exponential and these exact techniques will be intractable for larger NISQ devices \cite{paler2018nisq}. Thus, most recent approaches use heuristics to reduce the search space \cite{li2018tackling, paler2018nisq,CQCdepth}. This includes the IBM-QX contest-winning technique that is based on the A*-search algorithm \cite{zulehner2018efficient}. Ref.~\cite{ferrari2018demonstration} gives an approach for realising arbitrary parity-function oracles, subject to topological constraints, which are a special case of the family of maps we consider.
With most of these techniques, the size of the resulting circuit is very sensitive to the original placement of the logical qubits on the device \cite{paler2018influence}. Although algorithms have been proposed to find an optimal initial placement \textit{a priori} \cite{paler2018influence}, mapping techniques that have the freedom to build the initial placement while routing find a better initial state \cite{zulehner2018efficient}. Circuits can be routed even better if the initial mapping is adjusted based on the fully routed circuit \cite{li2018tackling}.

The authors of \cite{nash2019routing} have recently introduced a technique very similar to ours, which was developed independently. Both their article and a preprint of this paper appeared online within a few days of each other (our preprint~\cite{kissinger2019cnot} on 1/4/2019; \cite{nash2019routing} on 3/4/2019). The two approaches differ in that we give a slightly different strategy for preserving the upper triangular form based on Hamiltonian paths (which we then generalise to a recursive version based on a depth-first ordering). We also focus primarily on the case of CNOT circuits and sketch extensions to the case of CNOT+Rz and universal circuits using phase polynomials and the ZX-calculus, respectively. Ref.~\cite{nash2019routing} covers the CNOT and CNOT+Rz cases in detail, also relying on phase polynomials for the latter. We additionally optimise over initial qubit locations, which produces signficant improvements over the baseline compilers even for low CNOT counts (e.g. $>50\%$ improvement over Quilc with 8 CNOTs on IBM Q20). Finally, we compare performance relative to state of the art general-purpose circuit compilers, whereas \cite{nash2019routing} uses the CNOT circuit synthesis described in~\cite{PatelCNOT}, along with a na\"ive mapping procedure, as its baseline.

\section{Methods}\label{sec:methods}

\subsection{Background: Parity maps and Steiner trees}\label{sec:parity}
Our approach to CNOT mapping is based on re-synthesising the CNOT circuit from its corresponding parity map.
By a \textit{parity map}, we mean any reversible linear map on bitstrings. That is, we mean a bijective mapping from $N$-bitstrings to $N$-bitstrings where each bit in the output is a parity (i.e. XOR) of the input bits. It is a well-known fact that such maps exactly correspond to the action of CNOT circuits on computational basis states. It is therefore convenient to represent the action of a CNOT circuit on $N$ qubits as an $N \times N$ matrix over \bitfield.

If we consider an arbitrary such parity map, it is straightforward to check that post-composing a CNOT gate with a control on the $j$-th qubit and the target on the $k$-th qubit has the overall effect of adding the $j$-th row to the $k$-th row:
\[
\begin{tikzpicture}
	\begin{pgfonlayer}{nodelayer}
		\node [style=none] (0) at (-6.25, 1.75) {};
		\node [style=none] (1) at (-1.25, 1.75) {};
		\node [style=none] (2) at (-6.25, -1.75) {};
		\node [style=none] (3) at (-1.25, -1.75) {};
		\node [style=none] (4) at (-1.25, 0.75) {};
		\node [style=none] (5) at (-1.25, -0.25) {};
		\node [style=none] (6) at (-5.5, 2) {};
		\node [style=none] (7) at (-3.25, 2) {};
		\node [style=none] (8) at (-3.25, -2) {};
		\node [style=none] (9) at (-5.5, -2) {};
		\node [style=none] (10) at (-6.25, 0.75) {};
		\node [style=none] (11) at (-6.25, -0.25) {};
		\node [style=none] (12) at (-5.5, 1.75) {};
		\node [style=none] (13) at (-5.5, -1.75) {};
		\node [style=none] (14) at (-5.5, 0.75) {};
		\node [style=none] (15) at (-5.5, -0.25) {};
		\node [style=none] (16) at (-3.25, 1.75) {};
		\node [style=none] (17) at (-3.25, -1.75) {};
		\node [style=none] (18) at (-3.25, 0.75) {};
		\node [style=none] (19) at (-3.25, -0.25) {};
		\node [style=cnot ctrl] (20) at (-2.25, 0.75) {};
		\node [style=cnot targ] (21) at (-2.25, -0.25) {};
		\node [style=none] (22) at (-4.25, 0) {$P$};
		\node [style=none, rotate=90] (23) at (-1.75, -1) {...};
		\node [style=none, rotate=90] (24) at (-1.75, 1.25) {...};
		\node [style=none, rotate=90] (25) at (-1.75, 0.25) {...};
		\node [style=none, rotate=90] (26) at (-6, -1) {...};
		\node [style=none, rotate=90] (27) at (-6, 1.25) {...};
		\node [style=none, rotate=90] (28) at (-6, 0.25) {...};
		\node [style=none] (29) at (0.75, 1.75) {};
		\node [style=none] (30) at (4.5, 1.75) {};
		\node [style=none] (31) at (0.75, -1.75) {};
		\node [style=none] (32) at (4.5, -1.75) {};
		\node [style=none] (35) at (1.5, 2) {};
		\node [style=none] (36) at (3.75, 2) {};
		\node [style=none] (37) at (3.75, -2) {};
		\node [style=none] (38) at (1.5, -2) {};
		\node [style=none] (41) at (1.5, 1.75) {};
		\node [style=none] (42) at (1.5, -1.75) {};
		\node [style=none] (45) at (3.75, 1.75) {};
		\node [style=none] (46) at (3.75, -1.75) {};
		\node [style=none] (51) at (2.75, 0) {$P'$};
		\node [style=none, rotate=90] (53) at (4.25, 0) {...};
		\node [style=none, rotate=90] (56) at (1, 0) {...};
		\node [style=none] (57) at (0, 0) {$=$};
		\node [style=none] (58) at (-1, 0.75) {$i$};
		\node [style=none] (59) at (-1, -0.25) {$j$};
	\end{pgfonlayer}
	\begin{pgfonlayer}{edgelayer}
		\draw (6.center) to (7.center);
		\draw (7.center) to (8.center);
		\draw (8.center) to (9.center);
		\draw (9.center) to (6.center);
		\draw (0.center) to (12.center);
		\draw (10.center) to (14.center);
		\draw (11.center) to (15.center);
		\draw (2.center) to (13.center);
		\draw (16.center) to (1.center);
		\draw (18.center) to (4.center);
		\draw (19.center) to (5.center);
		\draw (17.center) to (3.center);
		\draw (20) to (21);
		\draw (35.center) to (36.center);
		\draw (36.center) to (37.center);
		\draw (37.center) to (38.center);
		\draw (38.center) to (35.center);
		\draw (29.center) to (41.center);
		\draw (31.center) to (42.center);
		\draw (45.center) to (30.center);
		\draw (46.center) to (32.center);
	\end{pgfonlayer}
\end{tikzpicture}
\quad\implies\quad
P \rowop{j}{i} P'
\]
Hence, there is an evident way to construct a CNOT circuit that realises an arbitrary parity matrix $P$. Simply perform Gauss-Jordan elimination on $P$, post-composing CNOTs for each primitive row operation. In the end, we will obtain $CP = 1$, where $C$ is a known CNOT circuit. Then, $P = C^{-1}$, where $C^{-1}$ is obtained from $C$ just by reversing the order of CNOT gates. In other words, in order to synthesise a CNOT circuit which realises parity map $P$, we simply perform Gauss-Jordan and store the primitive row operations used. Then the CNOT circuit corresponds exactly to that sequence of row operations, in reverse order. This technique, when combined with a simple heuristic for choosing appropriate row operations, is able to obtain asymptotically optimal CNOT realisations of a given parity map~\cite{PatelCNOT}.

In this paper, we modify the question: how can we construct a CNOT realisation of an arbitrary parity map if only certain CNOT gates are allowed? For example, suppose we have 9 qubits arranged in a $3 \times 3$ grid, and we only wish to allow CNOTs between nearest neighbours. Clearly this problem is equivalent to asking: how do we perform Gauss-Jordan elimination on a given \bitfield-matrix, when only certain primitive row operations are allowed?

Clearly, our only recourse is to use \textit{more} row operations in order to allow distant rows to essentially be added together via some intermediate steps. We present a strategy for doing this based on special kinds of spanning trees called Steiner trees.

Note that, by a \textit{graph} we always mean an indirected, simple graph with no self-loops, by a \textit{tree} we mean a graph with no cycles, and by a \textit{root tree} we mean a tree with a chosen vertex called the \textit{root}. For rooted trees, we use the standard terminology of \textit{parent} and \textit{child} to denote vertices adjacent to a given vertex which are closer to or farther from the root, respectively.

\begin{definition}\label{def:steiner-tree}
  For a graph $G$ and a subset $S$ of the vertices $V_G$ of $G$, a \textit{Steiner tree} $T$ is a minimal subgraph of $G$ that is furthermore a tree and has the property that $S \subseteq V_T$.
\end{definition}

Computing Steiner trees is NP-hard in general and the related decision problem is NP-complete~\cite{karp1972reducibility}. Indeed, the Steiner tree problem can be seen as a generalisation of the travelling salesperson problem, which allows an arbitrary tree to span a set of vertices rather than a single path, which can be seen as a tree with no branching. In practice, we will not need exactly optimal Steiner trees for our strategy to work, and many efficient heuristics exist for computing approximate Steiner trees~\cite{robins2000steiner,byrka2010steiner}. For our purposes, we use a very simple heuristic based on the Floyd-Warshall shortest-path algorith~\cite{cormen2009introduction} and minimal spanning trees.

A useful refinement of Steiner trees will be the following notion.

\begin{definition}\label{def:dec-steiner-tree}
  For a graph $G$, a total ordering $\leq$ of the vertices $V_G$, and a subset $S \subseteq V_G$, a \textit{decreasing Steiner tree} $T$ is a minimal rooted subtree of $G$ such that $S \subseteq V_T$ and every vertex in $T$ is larger its children with respect to $\leq$.
\end{definition}

\subsection{Constrained CNOT circuit extraction}\label{sec:algorithm}

\newcommand{\markspot}[2]{\tikz{\node[draw=#2,thick] {\color{#2} $#1$};}}
\newcommand{\sr}[1]{\markspot{#1}{red}}
\newcommand{\si}[1]{\markspot{#1}{green!50!black}}
\newcommand{\st}[1]{\markspot{#1}{blue}}

In this section, we will describe the algorithm \steinergauss, which performs Gauss-Jordan elimination of a parity matrix using only nearest-neighbour row operations for a given graph $G$. Consequently, this procedure can be used to synthesise a CNOT circuit implementing a given parity map using only nearest-neighbour CNOTs. The algorithm itself consists of two phases, \steinerdown and \steinerup, which respectively produce an upper triangular matrix and produce the identity matrix from an upper triangular matrix.

Initially, we will consider only graphs $G$ which have a Hamiltonian path, i.e. graphs $G$ which contain a connected path $P$ that visits each of the vertices in $G$ exactly once. For example, the $3 \times 3$ grid:
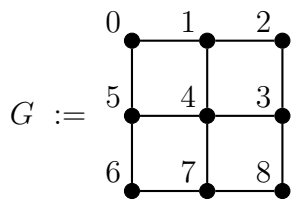
\[ G \ :=\  \begin{tikzpicture}
	\begin{pgfonlayer}{nodelayer}
		\node [style=black vertex] (0) at (-2, 2) {};
		\node [style=black vertex] (1) at (0, 2) {};
		\node [style=black vertex] (2) at (2, 2) {};
		\node [style=black vertex] (3) at (2, 0) {};
		\node [style=black vertex] (4) at (0, 0) {};
		\node [style=black vertex] (5) at (-2, 0) {};
		\node [style=black vertex] (6) at (-2, -2) {};
		\node [style=black vertex] (7) at (0, -2) {};
		\node [style=black vertex] (8) at (2, -2) {};
		\node [style=none] (9) at (-2.5, 2.5) {$0$};
		\node [style=none] (10) at (-0.5, 2.5) {$1$};
		\node [style=none] (11) at (1.5, 2.5) {$2$};
		\node [style=none] (12) at (-2.5, 0.5) {$5$};
		\node [style=none] (13) at (-0.5, 0.5) {$4$};
		\node [style=none] (14) at (1.5, 0.5) {$3$};
		\node [style=none] (15) at (-2.5, -1.5) {$6$};
		\node [style=none] (16) at (-0.5, -1.5) {$7$};
		\node [style=none] (17) at (1.5, -1.5) {$8$};
	\end{pgfonlayer}
	\begin{pgfonlayer}{edgelayer}
		\draw [style=black edge] (0) to (1);
		\draw [style=black edge] (1) to (2);
		\draw [style=black edge] (0) to (5);
		\draw [style=black edge] (5) to (6);
		\draw [style=black edge] (6) to (7);
		\draw [style=black edge] (7) to (8);
		\draw [style=black edge] (5) to (4);
		\draw [style=black edge] (4) to (3);
		\draw [style=black edge] (2) to (3);
		\draw [style=black edge] (3) to (8);
		\draw [style=black edge] (1) to (4);
		\draw [style=black edge] (4) to (7);
	\end{pgfonlayer}
\end{tikzpicture} \]
has a Hamiltonian path given by $[0, 1, 2, \ldots, 8]$. We will assume this path also provides a total ordering $\leq$ on the vertices of $G$. We will remove the assumption of a Hamiltonian path in the next section by providing a recursive algorithm capable of handling arbitrary graphs.

%
We begin by labelling the rows of our parity map $P$ by the vertices of the constraint graph $G$.
The first stage of our algorithm, \steinerdown, computes an upper triangular matrix. To do this, we wish to remove the non-zero elements below the diagonal. We do this one column at a time, starting with column $k := 0$ and proceeding left to right. Let $S$ be the set containing $k$ itself, as well as all of the vertices $j$ such that $j > k$ and $P_{jk} = 1$. That is, $S$ contains the diagonal element and all of the rows which contain 1s below the diagonal. For example, in the following parity map, $S = \{ 0, 2, 7 \}$:
\[
P = \scalebox{0.75}{$\left(
\begin{matrix}
  \sr1 & 0 & 1 & 1 & 1 & 1 & 0 & 0 & 1 \\ 
     0 & 1 & 1 & 0 & 1 & 1 & 1 & 1 & 0 \\ 
  \st1 & 0 & 0 & 0 & 1 & 1 & 1 & 0 & 1 \\ 
     0 & 1 & 0 & 0 & 0 & 0 & 0 & 0 & 0 \\ 
     0 & 1 & 1 & 1 & 1 & 0 & 1 & 1 & 1 \\ 
     0 & 0 & 0 & 0 & 1 & 0 & 1 & 0 & 0 \\ 
     0 & 0 & 1 & 0 & 0 & 1 & 0 & 0 & 1 \\ 
  \st1 & 1 & 1 & 1 & 0 & 0 & 1 & 1 & 0 \\ 
     0 & 0 & 1 & 0 & 0 & 1 & 0 & 1 & 1 \\ 
\end{matrix}
\right)
\begin{matrix}
\textrm{\footnotesize\color{gray} $0$} \\
\textrm{\footnotesize\color{gray} $1$} \\
\textrm{\footnotesize\color{gray} $2$} \\
\textrm{\footnotesize\color{gray} $3$} \\
\textrm{\footnotesize\color{gray} $4$} \\
\textrm{\footnotesize\color{gray} $5$} \\
\textrm{\footnotesize\color{gray} $6$} \\
\textrm{\footnotesize\color{gray} $7$} \\
\textrm{\footnotesize\color{gray} $8$} \\
\end{matrix}$}
\qquad\qquad
G \ :=\  \begin{tikzpicture}
	\begin{pgfonlayer}{nodelayer}
		\node [style=black vertex] (0) at (-2, 2) {};
		\node [style=black vertex] (1) at (0, 2) {};
		\node [style=black vertex] (2) at (2, 2) {};
		\node [style=black vertex] (3) at (2, 0) {};
		\node [style=black vertex] (4) at (0, 0) {};
		\node [style=black vertex] (5) at (-2, 0) {};
		\node [style=black vertex] (6) at (-2, -2) {};
		\node [style=black vertex] (7) at (0, -2) {};
		\node [style=black vertex] (8) at (2, -2) {};
		\node [style=none] (9) at (-2.5, 2.5) {$0$};
		\node [style=none] (10) at (-0.5, 2.5) {$1$};
		\node [style=none] (11) at (1.5, 2.5) {$2$};
		\node [style=none] (12) at (-2.5, 0.5) {$5$};
		\node [style=none] (13) at (-0.5, 0.5) {$4$};
		\node [style=none] (14) at (1.5, 0.5) {$3$};
		\node [style=none] (15) at (-2.5, -1.5) {$6$};
		\node [style=none] (16) at (-0.5, -1.5) {$7$};
		\node [style=none] (17) at (1.5, -1.5) {$8$};
		\node [style=sr circle] (18) at (-2, 2) {};
		\node [style=st circle] (19) at (2, 2) {};
		\node [style=st circle] (20) at (0, -2) {};
	\end{pgfonlayer}
	\begin{pgfonlayer}{edgelayer}
		\draw [style=black edge] (0) to (1);
		\draw [style=black edge] (1) to (2);
		\draw [style=black edge] (0) to (5);
		\draw [style=black edge] (5) to (6);
		\draw [style=black edge] (6) to (7);
		\draw [style=black edge] (7) to (8);
		\draw [style=black edge] (5) to (4);
		\draw [style=black edge] (4) to (3);
		\draw [style=black edge] (2) to (3);
		\draw [style=black edge] (3) to (8);
		\draw [style=black edge] (1) to (4);
		\draw [style=black edge] (4) to (7);
	\end{pgfonlayer}
\end{tikzpicture}
\]
These vertices are not adjacent in the graph, hence when we compute the Steiner tree $T$ containing $S$, we get some extra vertices, corresponding to rows that have 0s below the diagonal:
\begin{equation}\label{eq:33-mark}
P = \scalebox{0.75}{$\left(
\begin{matrix}
  \sr1 & 0 & 1 & 1 & 1 & 1 & 0 & 0 & 1 \\ 
  \si0 & 1 & 1 & 0 & 1 & 1 & 1 & 1 & 0 \\ 
  \st1 & 0 & 0 & 0 & 1 & 1 & 1 & 0 & 1 \\ 
     0 & 1 & 0 & 0 & 0 & 0 & 0 & 0 & 0 \\ 
  \si0 & 1 & 1 & 1 & 1 & 0 & 1 & 1 & 1 \\ 
     0 & 0 & 0 & 0 & 1 & 0 & 1 & 0 & 0 \\ 
     0 & 0 & 1 & 0 & 0 & 1 & 0 & 0 & 1 \\ 
  \st1 & 1 & 1 & 1 & 0 & 0 & 1 & 1 & 0 \\ 
     0 & 0 & 1 & 0 & 0 & 1 & 0 & 1 & 1 \\ 
\end{matrix}
\right)
\begin{matrix}
\textrm{\footnotesize\color{gray} $0$} \\
\textrm{\footnotesize\color{gray} $1$} \\
\textrm{\footnotesize\color{gray} $2$} \\
\textrm{\footnotesize\color{gray} $3$} \\
\textrm{\footnotesize\color{gray} $4$} \\
\textrm{\footnotesize\color{gray} $5$} \\
\textrm{\footnotesize\color{gray} $6$} \\
\textrm{\footnotesize\color{gray} $7$} \\
\textrm{\footnotesize\color{gray} $8$} \\
\end{matrix}$}
\qquad\qquad
G \ :=\  \begin{tikzpicture}
	\begin{pgfonlayer}{nodelayer}
		\node [style=black vertex] (0) at (-2, 2) {};
		\node [style=black vertex] (1) at (0, 2) {};
		\node [style=black vertex] (2) at (2, 2) {};
		\node [style=gray vertex] (3) at (2, 0) {};
		\node [style=black vertex] (4) at (0, 0) {};
		\node [style=gray vertex] (5) at (-2, 0) {};
		\node [style=gray vertex] (6) at (-2, -2) {};
		\node [style=black vertex] (7) at (0, -2) {};
		\node [style=gray vertex] (8) at (2, -2) {};
		\node [style=none] (9) at (-2.5, 2.5) {$0$};
		\node [style=none] (10) at (-0.5, 2.5) {$1$};
		\node [style=none] (11) at (1.5, 2.5) {$2$};
		\node [style=none] (12) at (-2.5, 0.5) {$5$};
		\node [style=none] (13) at (-0.5, 0.5) {$4$};
		\node [style=none] (14) at (1.5, 0.5) {$3$};
		\node [style=none] (15) at (-2.5, -1.5) {$6$};
		\node [style=none] (16) at (-0.5, -1.5) {$7$};
		\node [style=none] (17) at (1.5, -1.5) {$8$};
		\node [style=sr circle] (18) at (-2, 2) {};
		\node [style=st circle] (19) at (2, 2) {};
		\node [style=st circle] (20) at (0, -2) {};
		\node [style=si circle] (21) at (0, 2) {};
		\node [style=si circle] (22) at (0, 0) {};
	\end{pgfonlayer}
	\begin{pgfonlayer}{edgelayer}
		\draw [style=black edge] (0) to (1);
		\draw [style=black edge] (1) to (2);
		\draw [style=gray edge] (0) to (5);
		\draw [style=gray edge] (5) to (6);
		\draw [style=gray edge] (6) to (7);
		\draw [style=gray edge] (7) to (8);
		\draw [style=gray edge] (5) to (4);
		\draw [style=gray edge] (4) to (3);
		\draw [style=gray edge] (2) to (3);
		\draw [style=gray edge] (3) to (8);
		\draw [style=black edge] (1) to (4);
		\draw [style=black edge] (4) to (7);
	\end{pgfonlayer}
\end{tikzpicture}
\end{equation}
These extra vertices which need to be added to get a spanning tree are sometimes called \textit{Steiner points}. We consider the numbers in boxes above as decorating the corresponding vertices of the Steiner tree $T$. Initially there some 0s in the Steiner tree corresponding to Steiner points (and possibly the diagonal element), so we first `fill' the Steiner tree. That is, we add a row with a 1 to any neighbouring row in $T$ with a 0. Since the tree is connected, after finitely many iterations this will propagate 1s into every location in $T$:
\[
\scalebox{0.5}{$\left(
\begin{matrix}
  \sr1 & 0 & 1 & 1 & 1 & 1 & 0 & 0 & 1 \\ 
  \si0 & 1 & 1 & 0 & 1 & 1 & 1 & 1 & 0 \\ 
  \st1 & 0 & 0 & 0 & 1 & 1 & 1 & 0 & 1 \\ 
     0 & 1 & 0 & 0 & 0 & 0 & 0 & 0 & 0 \\ 
  \si0 & 1 & 1 & 1 & 1 & 0 & 1 & 1 & 1 \\ 
     0 & 0 & 0 & 0 & 1 & 0 & 1 & 0 & 0 \\ 
     0 & 0 & 1 & 0 & 0 & 1 & 0 & 0 & 1 \\ 
  \st1 & 1 & 1 & 1 & 0 & 0 & 1 & 1 & 0 \\ 
     0 & 0 & 1 & 0 & 0 & 1 & 0 & 1 & 1 \\ 
\end{matrix}
\right)$}
\rowop{1}{0}
\scalebox{0.5}{$\left(
\begin{matrix}
  \sr1 & 0 & 1 & 1 & 1 & 1 & 0 & 0 & 1 \\ 
  \si1 & 1 & 0 & 1 & 0 & 0 & 1 & 1 & 1 \\ 
  \st1 & 0 & 0 & 0 & 1 & 1 & 1 & 0 & 1 \\ 
     0 & 1 & 0 & 0 & 0 & 0 & 0 & 0 & 0 \\ 
  \si0 & 1 & 1 & 1 & 1 & 0 & 1 & 1 & 1 \\ 
     0 & 0 & 0 & 0 & 1 & 0 & 1 & 0 & 0 \\ 
     0 & 0 & 1 & 0 & 0 & 1 & 0 & 0 & 1 \\ 
  \st1 & 1 & 1 & 1 & 0 & 0 & 1 & 1 & 0 \\ 
     0 & 0 & 1 & 0 & 0 & 1 & 0 & 1 & 1 \\ 
\end{matrix}
\right)$}
\rowop{4}{7}
\scalebox{0.5}{$\left(
\begin{matrix}
  \sr1 & 0 & 1 & 1 & 1 & 1 & 0 & 0 & 1 \\ 
  \si1 & 1 & 0 & 1 & 0 & 0 & 1 & 1 & 1 \\ 
  \st1 & 0 & 0 & 0 & 1 & 1 & 1 & 0 & 1 \\ 
     0 & 1 & 0 & 0 & 0 & 0 & 0 & 0 & 0 \\ 
  \si1 & 0 & 0 & 0 & 1 & 0 & 0 & 0 & 1 \\ 
     0 & 0 & 0 & 0 & 1 & 0 & 1 & 0 & 0 \\ 
     0 & 0 & 1 & 0 & 0 & 1 & 0 & 0 & 1 \\ 
  \st1 & 1 & 1 & 1 & 0 & 0 & 1 & 1 & 0 \\ 
     0 & 0 & 1 & 0 & 0 & 1 & 0 & 1 & 1 \\ 
\end{matrix}
\right)$}
\]
After that, we can `empty' the Steiner tree by setting every location except for the diagonal to zero. We do this by regarding the diagonal as the root of the tree. For each leaf $v$ in $T$ with parent $w$, perform the row operation $R_v := R_v + R_w$, then remove $v$ from $T$. This terminates when there is only one vertex left in $T$, the root.
\[
\scalebox{0.5}{$\left(
\begin{matrix}
  \sr1 & 0 & 1 & 1 & 1 & 1 & 0 & 0 & 1 \\ 
  \si1 & 1 & 0 & 1 & 0 & 0 & 1 & 1 & 1 \\ 
  \st1 & 0 & 0 & 0 & 1 & 1 & 1 & 0 & 1 \\ 
     0 & 1 & 0 & 0 & 0 & 0 & 0 & 0 & 0 \\ 
  \si1 & 0 & 0 & 0 & 1 & 0 & 0 & 0 & 1 \\ 
     0 & 0 & 0 & 0 & 1 & 0 & 1 & 0 & 0 \\ 
     0 & 0 & 1 & 0 & 0 & 1 & 0 & 0 & 1 \\ 
  \st1 & 1 & 1 & 1 & 0 & 0 & 1 & 1 & 0 \\ 
     0 & 0 & 1 & 0 & 0 & 1 & 0 & 1 & 1 \\ 
\end{matrix}
\right)$}
\rowop{7}{4}
\scalebox{0.5}{$\left(
\begin{matrix}
  \sr1 & 0 & 1 & 1 & 1 & 1 & 0 & 0 & 1 \\ 
  \si1 & 1 & 0 & 1 & 0 & 0 & 1 & 1 & 1 \\ 
  \st1 & 0 & 0 & 0 & 1 & 1 & 1 & 0 & 1 \\ 
     0 & 1 & 0 & 0 & 0 & 0 & 0 & 0 & 0 \\ 
  \si1 & 0 & 0 & 0 & 1 & 0 & 0 & 0 & 1 \\ 
     0 & 0 & 0 & 0 & 1 & 0 & 1 & 0 & 0 \\ 
     0 & 0 & 1 & 0 & 0 & 1 & 0 & 0 & 1 \\ 
  \st0 & 1 & 1 & 1 & 1 & 0 & 1 & 1 & 1 \\ 
     0 & 0 & 1 & 0 & 0 & 1 & 0 & 1 & 1 \\ 
\end{matrix}
\right)$}
\rowop{2}{1}
\scalebox{0.5}{$\left(
\begin{matrix}
  \sr1 & 0 & 1 & 1 & 1 & 1 & 0 & 0 & 1 \\ 
  \si1 & 1 & 0 & 1 & 0 & 0 & 1 & 1 & 1 \\ 
  \st0 & 1 & 0 & 1 & 1 & 1 & 0 & 1 & 0 \\ 
     0 & 1 & 0 & 0 & 0 & 0 & 0 & 0 & 0 \\ 
  \si1 & 0 & 0 & 0 & 1 & 0 & 0 & 0 & 1 \\ 
     0 & 0 & 0 & 0 & 1 & 0 & 1 & 0 & 0 \\ 
     0 & 0 & 1 & 0 & 0 & 1 & 0 & 0 & 1 \\ 
  \st0 & 1 & 1 & 1 & 1 & 0 & 1 & 1 & 1 \\ 
     0 & 0 & 1 & 0 & 0 & 1 & 0 & 1 & 1 \\ 
\end{matrix}
\right)$}
\]
\[
\rowop{4}{1}
\scalebox{0.5}{$\left(
\begin{matrix}
  \sr1 & 0 & 1 & 1 & 1 & 1 & 0 & 0 & 1 \\ 
  \si1 & 1 & 0 & 1 & 0 & 0 & 1 & 1 & 1 \\ 
  \st0 & 1 & 0 & 1 & 1 & 1 & 0 & 1 & 0 \\ 
     0 & 1 & 0 & 0 & 0 & 0 & 0 & 0 & 0 \\ 
  \si0 & 1 & 0 & 1 & 1 & 0 & 1 & 1 & 0 \\ 
     0 & 0 & 0 & 0 & 1 & 0 & 1 & 0 & 0 \\ 
     0 & 0 & 1 & 0 & 0 & 1 & 0 & 0 & 1 \\ 
  \st0 & 1 & 1 & 1 & 1 & 0 & 1 & 1 & 1 \\ 
     0 & 0 & 1 & 0 & 0 & 1 & 0 & 1 & 1 \\ 
\end{matrix}
\right)$}
\rowop{1}{0}
\scalebox{0.5}{$\left(
\begin{matrix}
  \sr1 & 0 & 1 & 1 & 1 & 1 & 0 & 0 & 1 \\ 
  \si0 & 1 & 1 & 0 & 1 & 1 & 1 & 1 & 0 \\ 
  \st0 & 1 & 0 & 1 & 1 & 1 & 0 & 1 & 0 \\ 
     0 & 1 & 0 & 0 & 0 & 0 & 0 & 0 & 0 \\ 
  \si0 & 1 & 0 & 1 & 1 & 0 & 1 & 1 & 0 \\ 
     0 & 0 & 0 & 0 & 1 & 0 & 1 & 0 & 0 \\ 
     0 & 0 & 1 & 0 & 0 & 1 & 0 & 0 & 1 \\ 
  \st0 & 1 & 1 & 1 & 1 & 0 & 1 & 1 & 1 \\ 
     0 & 0 & 1 & 0 & 0 & 1 & 0 & 1 & 1 \\ 
\end{matrix}
\right)$}
\]
Since we only perform row operations along edges of $T$, which are a subset of the edges of $G$, the corresponding CNOTs in the circuit we synthesise will only be between neighbouring qubits. For example, the six row operations above (read from right to left) yield the following part of a CNOT circuit:
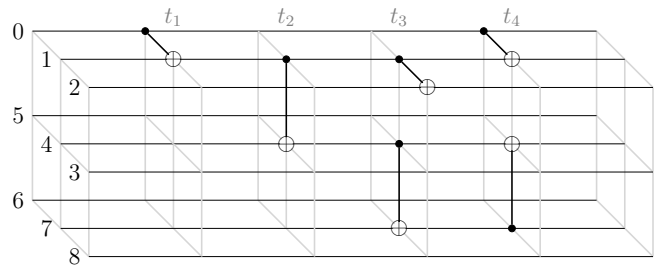
\[
\scalebox{0.75}{\begin{tikzpicture}
	\begin{pgfonlayer}{nodelayer}
		\node [style=none] (0) at (-10.5, 4) {};
		\node [style=none] (1) at (-9.5, 3) {};
		\node [style=none] (2) at (-8.5, 2) {};
		\node [style=none] (3) at (-10.5, 1) {};
		\node [style=none] (4) at (-9.5, 0) {};
		\node [style=none] (5) at (-8.5, -1) {};
		\node [style=none] (6) at (-10.5, -2) {};
		\node [style=none] (7) at (-9.5, -3) {};
		\node [style=none] (8) at (-8.5, -4) {};
		\node [style=none] (18) at (9.5, 4) {};
		\node [style=none] (19) at (10.5, 3) {};
		\node [style=none] (20) at (11.5, 2) {};
		\node [style=none] (21) at (9.5, 1) {};
		\node [style=none] (22) at (10.5, 0) {};
		\node [style=none] (23) at (11.5, -1) {};
		\node [style=none] (24) at (9.5, -2) {};
		\node [style=none] (25) at (10.5, -3) {};
		\node [style=none] (26) at (11.5, -4) {};
		\node [style=cnot ctrl] (27) at (-6.5, 4) {};
		\node [style=cnot targ] (28) at (-5.5, 3) {};
		\node [style=none] (29) at (-4.5, 2) {};
		\node [style=none] (30) at (-6.5, 1) {};
		\node [style=none] (32) at (-4.5, -1) {};
		\node [style=none] (33) at (-6.5, -2) {};
		\node [style=none] (34) at (-5.5, -3) {};
		\node [style=none] (35) at (-4.5, -4) {};
		\node [style=none] (36) at (-2.5, 4) {};
		\node [style=cnot ctrl] (37) at (-1.5, 3) {};
		\node [style=none] (38) at (-0.5, 2) {};
		\node [style=none] (39) at (-2.5, 1) {};
		\node [style=cnot targ] (40) at (-1.5, 0) {};
		\node [style=none] (41) at (-0.5, -1) {};
		\node [style=none] (42) at (-2.5, -2) {};
		\node [style=none] (43) at (-1.5, -3) {};
		\node [style=none] (44) at (-0.5, -4) {};
		\node [style=none] (45) at (1.5, 4) {};
		\node [style=cnot ctrl] (46) at (2.5, 3) {};
		\node [style=cnot targ] (47) at (3.5, 2) {};
		\node [style=none] (48) at (1.5, 1) {};
		\node [style=cnot ctrl] (49) at (2.5, 0) {};
		\node [style=none] (50) at (3.5, -1) {};
		\node [style=none] (51) at (1.5, -2) {};
		\node [style=cnot targ] (52) at (2.5, -3) {};
		\node [style=none] (53) at (3.5, -4) {};
		\node [style=cnot ctrl] (54) at (5.5, 4) {};
		\node [style=cnot targ] (55) at (6.5, 3) {};
		\node [style=none] (56) at (7.5, 2) {};
		\node [style=none] (57) at (5.5, 1) {};
		\node [style=cnot targ] (58) at (6.5, 0) {};
		\node [style=none] (59) at (7.5, -1) {};
		\node [style=none] (60) at (5.5, -2) {};
		\node [style=cnot ctrl] (61) at (6.5, -3) {};
		\node [style=none] (62) at (7.5, -4) {};
		\node [style=none] (63) at (-11, 4) {$0$};
		\node [style=none] (64) at (-10, 0) {$4$};
		\node [style=none] (65) at (-10, 3) {$1$};
		\node [style=none] (66) at (-9, 2) {$2$};
		\node [style=none] (67) at (-11, -2) {$6$};
		\node [style=none] (68) at (-10, -3) {$7$};
		\node [style=none] (69) at (-9, -4) {$8$};
		\node [style=none] (70) at (-11, 1) {$5$};
		\node [style=none] (72) at (-9, -1) {$3$};
		\node [style=none] (73) at (-5.5, 0) {};
		\node [style=none] (74) at (-5.5, 4.5) {\color{gray}$t_1$};
		\node [style=none] (75) at (-1.5, 4.5) {\color{gray}$t_2$};
		\node [style=none] (76) at (2.5, 4.5) {\color{gray}$t_3$};
		\node [style=none] (77) at (6.5, 4.5) {\color{gray}$t_4$};
	\end{pgfonlayer}
	\begin{pgfonlayer}{edgelayer}
		\draw [style=gray edge] (46) to (49);
		\draw [style=gray edge] (55) to (58);
		\draw [style=gray edge] (40) to (43.center);
		\draw [style=gray edge] (28) to (73.center);
		\draw [style=gray edge] (73.center) to (34.center);
		\draw [style=gray edge] (54) to (57.center);
		\draw [style=gray edge] (57.center) to (60.center);
		\draw [style=gray edge] (45) to (48.center);
		\draw [style=gray edge] (48.center) to (51.center);
		\draw [style=gray edge] (51.center) to (52);
		\draw [style=gray edge] (52) to (53.center);
		\draw [style=gray edge] (60.center) to (61);
		\draw [style=gray edge] (61) to (62.center);
		\draw [style=gray edge] (22.center) to (25.center);
		\draw [style=gray edge] (18.center) to (21.center);
		\draw [style=gray edge] (21.center) to (24.center);
		\draw [style=gray edge] (24.center) to (25.center);
		\draw [style=gray edge] (25.center) to (26.center);
		\draw [style=gray edge] (27) to (30.center);
		\draw [style=gray edge] (30.center) to (33.center);
		\draw [style=gray edge] (33.center) to (34.center);
		\draw [style=gray edge] (34.center) to (35.center);
		\draw [style=gray edge] (36.center) to (39.center);
		\draw [style=gray edge] (39.center) to (42.center);
		\draw [style=gray edge] (42.center) to (43.center);
		\draw [style=gray edge] (43.center) to (44.center);
		\draw [style=gray edge] (3.center) to (4.center);
		\draw [style=gray edge] (4.center) to (5.center);
		\draw [style=gray edge] (1.center) to (4.center);
		\draw [style=gray edge] (4.center) to (7.center);
		\draw [style=gray edge] (2.center) to (5.center);
		\draw [style=gray edge] (5.center) to (8.center);
		\draw [style=gray edge] (0.center) to (3.center);
		\draw [style=gray edge] (3.center) to (6.center);
		\draw [style=gray edge] (6.center) to (7.center);
		\draw [style=gray edge] (7.center) to (8.center);
		\draw [style=gray edge] (0.center) to (1.center);
		\draw [style=gray edge] (1.center) to (2.center);
		\draw [style=gray edge] (21.center) to (22.center);
		\draw [style=gray edge] (22.center) to (23.center);
		\draw [style=gray edge] (19.center) to (22.center);
		\draw [style=gray edge] (20.center) to (23.center);
		\draw [style=gray edge] (23.center) to (26.center);
		\draw [style=gray edge] (18.center) to (19.center);
		\draw [style=gray edge] (19.center) to (20.center);
		\draw (0.center) to (18.center);
		\draw (1.center) to (19.center);
		\draw (2.center) to (20.center);
		\draw (3.center) to (21.center);
		\draw (4.center) to (22.center);
		\draw (5.center) to (23.center);
		\draw (6.center) to (24.center);
		\draw (7.center) to (25.center);
		\draw (8.center) to (26.center);
		\draw [style=gray edge] (29.center) to (32.center);
		\draw [style=gray edge] (32.center) to (35.center);
		\draw [style=black edge] (27) to (28);
		\draw [style=gray edge] (28) to (29.center);
		\draw [style=black edge] (37) to (40);
		\draw [style=gray edge] (38.center) to (41.center);
		\draw [style=gray edge] (41.center) to (44.center);
		\draw [style=gray edge] (36.center) to (37);
		\draw [style=gray edge] (37) to (38.center);
		\draw [style=black edge] (49) to (52);
		\draw [style=gray edge] (47.center) to (50.center);
		\draw [style=gray edge] (50.center) to (53.center);
		\draw [style=gray edge] (45.center) to (46);
		\draw [style=black edge] (46) to (47);
		\draw [style=black edge] (58) to (61);
		\draw [style=gray edge] (56.center) to (59.center);
		\draw [style=gray edge] (59.center) to (62.center);
		\draw [style=black edge] (54) to (55);
		\draw [style=gray edge] (55) to (56.center);
		\draw [style=gray edge] (30.center) to (73.center);
		\draw [style=gray edge] (73.center) to (32.center);
		\draw [style=gray edge] (40) to (41.center);
		\draw [style=gray edge] (39.center) to (40);
		\draw [style=gray edge] (48.center) to (49);
		\draw [style=gray edge] (49) to (50.center);
		\draw [style=gray edge] (57.center) to (58);
		\draw [style=gray edge] (58) to (59.center);
	\end{pgfonlayer}
\end{tikzpicture}}
\]
Note that in this phase, we have some freedom to choose which row operations to perform. Here we have taken a greedy strategy for maximising the number of row operations that can be done in parallel.

Having put the first column in upper triangular form, we delete the corresponding root vertice from $G$ and then proceed to the next column, building up our CNOT circuit from right-to-left. Since we proceed in order along a Hamiltonian path, the graph never becomes disconnected and always has a Hamiltonian path. Hence it is always possible to find a Steiner tree for any combination of points. \steinerdown terminates after we remove the last vertex from $G$ with a matrix in upper triangular form.

The second stage of our algorithm,
\steinerup, starts with the original graph $G$ and a parity map $P$ in upper triangular form and removes any 1s that are above the diagonal, yielding the reduced echelon form (which in our case is always an identity matrix). It works in almost the same way, except that some care must be taken not to destroy the existing upper triangular structure. To do this, we must always perform \textit{decreasing} row operations. That is, we must perform operations of the form $R_j := R_j + R_i$ only when $j < i$. This is where Definition~\ref{def:dec-steiner-tree} comes in. Starting with the last column, let the set $S$ consist of the diagonal element and the rows which contain 1s above the diagonal. We then compute a decreasing Steiner tree for $S$ whose root is the diagonal element. This is always possible, because in the worse case we can just take the Hamiltonian path for $T$, but in general we can take shortcuts. For example:
\[
P = \scalebox{0.5}{$\left(
\begin{matrix}
1 & 0 & 1 & 1 & 1 & 1 & 0 & 0 &    0 \\
0 & 1 & 1 & 0 & 1 & 1 & 1 & 1 &    0 \\
0 & 0 & 1 & 1 & 0 & 0 & 1 & 0 & \st1 \\
0 & 0 & 0 & 1 & 1 & 1 & 0 & 1 & \si0 \\
0 & 0 & 0 & 0 & 1 & 0 & 0 & 1 & \st1 \\
0 & 0 & 0 & 0 & 0 & 1 & 0 & 1 &    0 \\
0 & 0 & 0 & 0 & 0 & 0 & 1 & 0 &    0 \\
0 & 0 & 0 & 0 & 0 & 0 & 0 & 1 & \si0 \\
0 & 0 & 0 & 0 & 0 & 0 & 0 & 0 & \sr1 \\
\end{matrix}\right)
\begin{matrix}
\textrm{\footnotesize\color{gray} $0$} \\
\textrm{\footnotesize\color{gray} $1$} \\
\textrm{\footnotesize\color{gray} $2$} \\
\textrm{\footnotesize\color{gray} $3$} \\
\textrm{\footnotesize\color{gray} $4$} \\
\textrm{\footnotesize\color{gray} $5$} \\
\textrm{\footnotesize\color{gray} $6$} \\
\textrm{\footnotesize\color{gray} $7$} \\
\textrm{\footnotesize\color{gray} $8$} \\
\end{matrix}$}
\qquad\qquad
G \ :=\  \begin{tikzpicture}
	\begin{pgfonlayer}{nodelayer}
		\node [style=gray vertex] (0) at (-2, 2) {};
		\node [style=gray vertex] (1) at (0, 2) {};
		\node [style=black vertex] (2) at (2, 2) {};
		\node [style=black vertex] (3) at (2, 0) {};
		\node [style=black vertex] (4) at (0, 0) {};
		\node [style=gray vertex] (5) at (-2, 0) {};
		\node [style=gray vertex] (6) at (-2, -2) {};
		\node [style=black vertex] (7) at (0, -2) {};
		\node [style=black vertex] (8) at (2, -2) {};
		\node [style=none] (9) at (-2.5, 2.5) {$0$};
		\node [style=none] (10) at (-0.5, 2.5) {$1$};
		\node [style=none] (11) at (1.5, 2.5) {$2$};
		\node [style=none] (12) at (-2.5, 0.5) {$5$};
		\node [style=none] (13) at (-0.5, 0.5) {$4$};
		\node [style=none] (14) at (1.5, 0.5) {$3$};
		\node [style=none] (15) at (-2.5, -1.5) {$6$};
		\node [style=none] (16) at (-0.5, -1.5) {$7$};
		\node [style=none] (17) at (1.5, -1.5) {$8$};
		\node [style=st circle] (19) at (2, 2) {};
		\node [style=st circle] (20) at (0, 0) {};
		\node [style=si circle] (21) at (2, 0) {};
		\node [style=si circle] (22) at (0, -2) {};
		\node [style=sr circle] (23) at (2, -2) {};
	\end{pgfonlayer}
	\begin{pgfonlayer}{edgelayer}
		\draw [style=gray edge] (0) to (1);
		\draw [style=gray edge] (1) to (2);
		\draw [style=gray edge] (0) to (5);
		\draw [style=gray edge] (5) to (6);
		\draw [style=gray edge] (6) to (7);
		\draw [style=black edge] (7) to (8);
		\draw [style=gray edge] (5) to (4);
		\draw [style=gray edge] (4) to (3);
		\draw [style=black edge] (2) to (3);
		\draw [style=black edge] (3) to (8);
		\draw [style=gray edge] (1) to (4);
		\draw [style=black edge] (4) to (7);
	\end{pgfonlayer}
\end{tikzpicture}
\]
Here, we have $S = \{ 2, 4, 8 \}$ and $T$ has vertices $\{ 2, 3, 4, 7, 8 \}$. Note there is indeed a smaller Steiner tree which does not contain vertex $7$, but in that case $4$ would need to be a child of $3$, hence it is not a decreasing Steiner tree.

Once we have a decreasing Steiner tree, we can `fill' the tree with 1s by decreasing row operations. This is always possible since at this stage, the root always contains a 1 and all edges from parents to children are decreasing. We can then `empty' the tree again just as we did in \steinerdown, noting that every row operation at this stage is applied from a parent to its child (and hence is decreasing). The resulting row operations in the above example are thus:
\[
\scalebox{0.5}{$\left(
\begin{matrix}
1 & 0 & 1 & 1 & 1 & 1 & 0 & 0 &    0 \\
0 & 1 & 1 & 0 & 1 & 1 & 1 & 1 &    0 \\
0 & 0 & 1 & 1 & 0 & 0 & 1 & 0 & \st1 \\
0 & 0 & 0 & 1 & 1 & 1 & 0 & 1 & \si0 \\
0 & 0 & 0 & 0 & 1 & 0 & 0 & 1 & \st1 \\
0 & 0 & 0 & 0 & 0 & 1 & 0 & 1 &    0 \\
0 & 0 & 0 & 0 & 0 & 0 & 1 & 0 &    0 \\
0 & 0 & 0 & 0 & 0 & 0 & 0 & 1 & \si0 \\
0 & 0 & 0 & 0 & 0 & 0 & 0 & 0 & \sr1 \\
\end{matrix}\right)$}
\rowop{3}{8}
\rowop{7}{8}
\rowop{2}{3}
\ \ \cdots
\]
\[
\cdots \ \ 
\rowop{4}{7}
\rowop{7}{8}
\rowop{3}{8}
\scalebox{0.5}{$\left(\begin{matrix}
1 & 0 & 1 & 1 & 1 & 1 & 0 & 0 &    0 \\
0 & 1 & 1 & 0 & 1 & 1 & 1 & 1 &    0 \\
0 & 0 & 1 & 0 & 1 & 1 & 1 & 1 & \st0 \\
0 & 0 & 0 & 1 & 1 & 1 & 0 & 1 & \si0 \\
0 & 0 & 0 & 0 & 1 & 0 & 0 & 0 & \st0 \\
0 & 0 & 0 & 0 & 0 & 1 & 0 & 1 &    0 \\
0 & 0 & 0 & 0 & 0 & 0 & 1 & 0 &    0 \\
0 & 0 & 0 & 0 & 0 & 0 & 0 & 1 & \si0 \\
0 & 0 & 0 & 0 & 0 & 0 & 0 & 0 & \sr1 \\
\end{matrix}\right)$}
\]
We can therefore prepend the following section to our CNOT circuit:
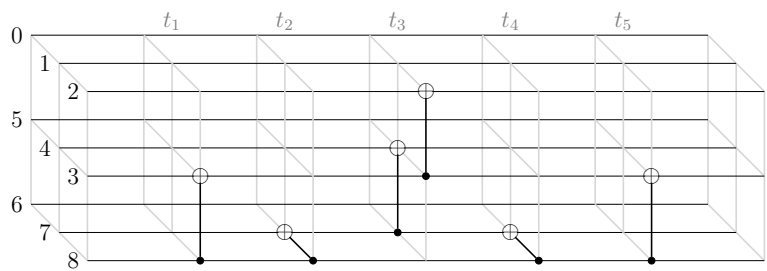
\[
\scalebox{0.75}{\begin{tikzpicture}
	\begin{pgfonlayer}{nodelayer}
		\node [style=none] (0) at (-10, 4) {};
		\node [style=none] (1) at (-9, 3) {};
		\node [style=none] (2) at (-8, 2) {};
		\node [style=none] (3) at (-10, 1) {};
		\node [style=none] (4) at (-9, 0) {};
		\node [style=none] (5) at (-8, -1) {};
		\node [style=none] (6) at (-10, -2) {};
		\node [style=none] (7) at (-9, -3) {};
		\node [style=none] (8) at (-8, -4) {};
		\node [style=none] (18) at (14, 4) {};
		\node [style=none] (19) at (15, 3) {};
		\node [style=none] (20) at (16, 2) {};
		\node [style=none] (21) at (14, 1) {};
		\node [style=none] (22) at (15, 0) {};
		\node [style=none] (23) at (16, -1) {};
		\node [style=none] (24) at (14, -2) {};
		\node [style=none] (25) at (15, -3) {};
		\node [style=none] (26) at (16, -4) {};
		\node [style=none] (27) at (-6, 4) {};
		\node [style=none] (28) at (-5, 3) {};
		\node [style=none] (29) at (-4, 2) {};
		\node [style=none] (30) at (-6, 1) {};
		\node [style=cnot targ] (32) at (-4, -1) {};
		\node [style=none] (33) at (-6, -2) {};
		\node [style=none] (34) at (-5, -3) {};
		\node [style=cnot ctrl] (35) at (-4, -4) {};
		\node [style=none] (36) at (-2, 4) {};
		\node [style=none] (37) at (-1, 3) {};
		\node [style=none] (38) at (0, 2) {};
		\node [style=none] (39) at (-2, 1) {};
		\node [style=none] (40) at (-1, 0) {};
		\node [style=none] (41) at (0, -1) {};
		\node [style=none] (42) at (-2, -2) {};
		\node [style=cnot targ] (43) at (-1, -3) {};
		\node [style=cnot ctrl] (44) at (0, -4) {};
		\node [style=none] (45) at (2, 4) {};
		\node [style=none] (46) at (3, 3) {};
		\node [style=cnot targ] (47) at (4, 2) {};
		\node [style=none] (48) at (2, 1) {};
		\node [style=cnot targ] (49) at (3, 0) {};
		\node [style=cnot ctrl] (50) at (4, -1) {};
		\node [style=none] (51) at (2, -2) {};
		\node [style=cnot ctrl] (52) at (3, -3) {};
		\node [style=none] (53) at (4, -4) {};
		\node [style=none] (54) at (6, 4) {};
		\node [style=none] (55) at (7, 3) {};
		\node [style=none] (56) at (8, 2) {};
		\node [style=none] (57) at (6, 1) {};
		\node [style=none] (58) at (7, 0) {};
		\node [style=none] (59) at (8, -1) {};
		\node [style=none] (60) at (6, -2) {};
		\node [style=cnot targ] (61) at (7, -3) {};
		\node [style=cnot ctrl] (62) at (8, -4) {};
		\node [style=none] (63) at (-10.5, 4) {$0$};
		\node [style=none] (64) at (-9.5, 0) {$4$};
		\node [style=none] (65) at (-9.5, 3) {$1$};
		\node [style=none] (66) at (-8.5, 2) {$2$};
		\node [style=none] (67) at (-10.5, -2) {$6$};
		\node [style=none] (68) at (-9.5, -3) {$7$};
		\node [style=none] (69) at (-8.5, -4) {$8$};
		\node [style=none] (70) at (-10.5, 1) {$5$};
		\node [style=none] (72) at (-8.5, -1) {$3$};
		\node [style=none] (73) at (-5, 0) {};
		\node [style=none] (74) at (-5, 4.5) {\color{gray} $t_1$};
		\node [style=none] (75) at (-1, 4.5) {\color{gray} $t_2$};
		\node [style=none] (76) at (3, 4.5) {\color{gray} $t_3$};
		\node [style=none] (77) at (7, 4.5) {\color{gray} $t_4$};
		\node [style=none] (78) at (10, 4) {};
		\node [style=none] (79) at (11, 3) {};
		\node [style=none] (80) at (12, 2) {};
		\node [style=none] (81) at (10, 1) {};
		\node [style=none] (82) at (11, 0) {};
		\node [style=cnot targ] (83) at (12, -1) {};
		\node [style=none] (84) at (10, -2) {};
		\node [style=none] (85) at (11, -3) {};
		\node [style=cnot ctrl] (86) at (12, -4) {};
		\node [style=none] (87) at (11, 4.5) {\color{gray} $t_5$};
	\end{pgfonlayer}
	\begin{pgfonlayer}{edgelayer}
		\draw [style=gray edge] (58.center) to (61);
		\draw [style=gray edge] (46.center) to (49);
		\draw [style=gray edge] (55.center) to (58.center);
		\draw [style=gray edge] (40.center) to (43);
		\draw [style=gray edge] (28.center) to (73.center);
		\draw [style=gray edge] (73.center) to (34.center);
		\draw [style=gray edge] (54.center) to (57.center);
		\draw [style=gray edge] (57.center) to (60.center);
		\draw [style=gray edge] (45.center) to (48.center);
		\draw [style=gray edge] (48.center) to (51.center);
		\draw [style=gray edge] (51.center) to (52);
		\draw [style=gray edge] (52) to (53.center);
		\draw [style=gray edge] (60.center) to (61);
		\draw [style=black edge] (61) to (62);
		\draw [style=gray edge] (22.center) to (25.center);
		\draw [style=gray edge] (18.center) to (21.center);
		\draw [style=gray edge] (21.center) to (24.center);
		\draw [style=gray edge] (24.center) to (25.center);
		\draw [style=gray edge] (25.center) to (26.center);
		\draw [style=gray edge] (27.center) to (30.center);
		\draw [style=gray edge] (30.center) to (33.center);
		\draw [style=gray edge] (33.center) to (34.center);
		\draw [style=gray edge] (34.center) to (35);
		\draw [style=gray edge] (36.center) to (39.center);
		\draw [style=gray edge] (39.center) to (42.center);
		\draw [style=gray edge] (42.center) to (43);
		\draw [style=black edge] (43) to (44);
		\draw [style=gray edge] (3.center) to (4.center);
		\draw [style=gray edge] (4.center) to (5.center);
		\draw [style=gray edge] (1.center) to (4.center);
		\draw [style=gray edge] (4.center) to (7.center);
		\draw [style=gray edge] (2.center) to (5.center);
		\draw [style=gray edge] (5.center) to (8.center);
		\draw [style=gray edge] (0.center) to (3.center);
		\draw [style=gray edge] (3.center) to (6.center);
		\draw [style=gray edge] (6.center) to (7.center);
		\draw [style=gray edge] (7.center) to (8.center);
		\draw [style=gray edge] (0.center) to (1.center);
		\draw [style=gray edge] (1.center) to (2.center);
		\draw [style=gray edge] (21.center) to (22.center);
		\draw [style=gray edge] (22.center) to (23.center);
		\draw [style=gray edge] (19.center) to (22.center);
		\draw [style=gray edge] (20.center) to (23.center);
		\draw [style=gray edge] (23.center) to (26.center);
		\draw [style=gray edge] (18.center) to (19.center);
		\draw [style=gray edge] (19.center) to (20.center);
		\draw (0.center) to (18.center);
		\draw (1.center) to (19.center);
		\draw (2.center) to (20.center);
		\draw (3.center) to (21.center);
		\draw (4.center) to (22.center);
		\draw (5.center) to (23.center);
		\draw (6.center) to (24.center);
		\draw (7.center) to (25.center);
		\draw (8.center) to (26.center);
		\draw [style=gray edge] (29.center) to (32);
		\draw [style=black edge] (32) to (35);
		\draw [style=gray edge] (27.center) to (28.center);
		\draw [style=gray edge] (28.center) to (29.center);
		\draw [style=gray edge] (37.center) to (40.center);
		\draw [style=gray edge] (38.center) to (41.center);
		\draw [style=gray edge] (41.center) to (44);
		\draw [style=gray edge] (36.center) to (37.center);
		\draw [style=gray edge] (37.center) to (38.center);
		\draw [style=black edge] (49) to (52);
		\draw [style=black edge] (47) to (50);
		\draw [style=gray edge] (50) to (53.center);
		\draw [style=gray edge] (45.center) to (46.center);
		\draw [style=gray edge] (46.center) to (47);
		\draw [style=gray edge] (56.center) to (59.center);
		\draw [style=gray edge] (59.center) to (62);
		\draw [style=gray edge] (54.center) to (55.center);
		\draw [style=gray edge] (55.center) to (56.center);
		\draw [style=gray edge] (30.center) to (73.center);
		\draw [style=gray edge] (73.center) to (32);
		\draw [style=gray edge] (40.center) to (41.center);
		\draw [style=gray edge] (39.center) to (40.center);
		\draw [style=gray edge] (48.center) to (49);
		\draw [style=gray edge] (49) to (50);
		\draw [style=gray edge] (57.center) to (58.center);
		\draw [style=gray edge] (58.center) to (59.center);
		\draw [style=gray edge] (82.center) to (85.center);
		\draw [style=gray edge] (78.center) to (81.center);
		\draw [style=gray edge] (81.center) to (84.center);
		\draw [style=gray edge] (84.center) to (85.center);
		\draw [style=gray edge] (85.center) to (86);
		\draw [style=gray edge] (81.center) to (82.center);
		\draw [style=gray edge] (82.center) to (83);
		\draw [style=gray edge] (79.center) to (82.center);
		\draw [style=gray edge] (80.center) to (83);
		\draw [style=black edge] (83) to (86);
		\draw [style=gray edge] (78.center) to (79.center);
		\draw [style=gray edge] (79.center) to (80.center);
	\end{pgfonlayer}
\end{tikzpicture}}
\]

Once a column is done, we can delete it from $G$ and take the next highest column. Again, since we traverse along a Hamiltonian path (backwards this time), the graph $G$ never becomes disconnected and always has a Hamiltonian path. \steinerup then terminates with $P$ equal to the identity matrix. The corresponding CNOT circuit then implements the original parity map using only nearest-neighbour CNOT gates.

Much like other approaches, the size of the final circuit is very sensitive to the initial layout of the logical qubits within the physical architecture, especially for relatively small CNOT circuits. For us, a good choice of qubit positions means smaller Steiner trees, which in turn means that fewer extra CNOTs are added. The placement of qubits on the architecture is equivalent to permuting the rows and columns of the parity map before applying the algorithm described above. Hence, we use a genetic algorithm \cite{goldberg1989genetic} to find an optimal permutation such that the resulting circuit contains as little gates as possible.

\newcommand{\bneg}{\bf\color{green!60!black}}
\newcommand{\eneg}{\rm\color{gray}}

\begin{table}
  \centering
  
  \scalebox{0.8}{
  \!\!\!\!\begin{tabular}{l|c|rl|rl|rl|r|r}
  \textbf{Architecture} &
  \textbf{\#} &
  \multicolumn{2}{|c|}{\textbf{QuilC}} &
  \multicolumn{2}{|c|}{\textbf{\textsf{t$|$ket$\rangle$}}} &
  \multicolumn{2}{|c|}{\textbf{Steiner}} &
  $<$\textbf{QuilC} &
  $<$\textbf{\textsf{t$|$ket$\rangle$}}
  \\
  \hline
      \begin{filecontents*}{data/steiner_data_tex.csv}
arch,ngates,quil,quilo,tket,tketo,steiner,steinero,pquil,ptket
9q-square,3,3.8,27,3.6,20,3,\bneg{}0\eneg{},21,17
9q-square,5,10.82,116,6.4,28,5.2,4,52,19
9q-square,10,20.08,101,16.95,70,11.6,16,42,32
9q-square,20,46.24,131,40.75,104,25.85,29,44,37
9q-square,30,72.89,143,66.15,121,35.55,19,51,46
16q-square,4,6.14,54,5.8,45,4.44,11,28,23
16q-square,8,19.68,146,12.95,62,12.41,55,37,4
16q-square,16,48.13,201,36.2,126,33.08,107,31,9
16q-square,32,106.75,234,94.45,195,82.95,159,22,12
16q-square,64,225.69,253,203.75,218,147.38,130,35,28
16q-square,128,457.35,257,436.25,241,168.12,31,63,61
16q-square,256,925.85,262,922.65,260,169.28,\bneg{}-34\eneg{},82,82
rigetti-16q-aspen,4,7.05,76,7.15,79,4.15,4,41,42
rigetti-16q-aspen,8,28.2,253,17.2,115,11.22,40,60,35
rigetti-16q-aspen,16,69.15,332,52,225,33.95,112,51,35
rigetti-16q-aspen,32,147.3,360,144.95,353,101.75,218,31,29
rigetti-16q-aspen,64,324.6,407,322.85,404,189.15,196,42,41
rigetti-16q-aspen,128,664.65,419,666.15,420,220.75,72,67,66
rigetti-16q-aspen,256,1367.89,434,1361.6,432,222.15,\bneg{}-13\eneg{},84,83
ibm-qx5,4,6.75,69,4.3,8,4,\bneg{}0\eneg{},41,7
ibm-qx5,8,23.7,196,14.75,84,8.95,12,62,39
ibm-qx5,16,60.5,278,47.5,197,26.55,66,56,44
ibm-qx5,32,140.05,338,122.95,284,84.4,164,40,31
ibm-qx5,64,301.05,370,278.7,335,152.65,139,49,45
ibm-qx5,128,600.9,369,597.65,367,188.25,47,69,69
ibm-qx5,256,1247.8,387,1258.8,392,193.8,\bneg{}-24\eneg{},84,85
ibm-q20-tokyo,4,5.5,38,6.05,51,4,\bneg{}0\eneg{},27,34
ibm-q20-tokyo,8,17.3,116,12,50,7.69,\bneg{}-4\eneg{},56,36
ibm-q20-tokyo,16,43.83,174,29.05,82,20.44,28,53,30
ibm-q20-tokyo,32,93.58,192,78.15,144,66.93,109,28,14
ibm-q20-tokyo,64,215.9,237,181.25,183,165.6,159,23,9
ibm-q20-tokyo,128,432.65,238,391.85,206,237.64,86,45,39
ibm-q20-tokyo,256,860.74,236,789.3,208,245.84,\bneg{}-4\eneg{},71,69
\end{filecontents*}
\csvreader[late after line=\\, head to column names]
      {data/steiner_data_tex.csv} 
      {} 
      {\texttt{\arch} & \ngates & \quil & {\color{gray} (\quilo\%)} & \tket & {\color{gray} (\tketo\%)} & \steiner &
      {\color{gray} (\steinero\%)} & \pquil\% & \ptket\%} 
   \end{tabular}}
   \caption{The average number of CNOTs needed to map random circuits containing `\#' CNOT gates. The first column shows the architecture mapped to and the second column the original number of CNOT gates. The remaining columns show the average 2-qubit gate count after mapping 20 random circuits and percent improvement in total CNOT count of our approach vs. QuilC and \textsf{t$|$ket$\rangle$}. The percentages in parentheses show mapping overheads (i.e. percentage of the original gate count that was added during mapping), where negative overheads indicate a mapped circuit that is smaller than the original.}\label{table:results}
\end{table}

\section{Results}\label{sec:results}


We work with a fixed set of randomly-generated CNOT circuits on 9, 16, and 20 qubits.
The 9-qubit CNOT circuits have either 3, 5, 10, 20 or 30 gates, whereas the other circuits have either 4, 8, 16, 32, 64, 128 or 256 gates. For each of these gate counts, we generated 20 different random circuits, yielding a test set of 380 random circuits.

We compared our algorithm to the QuilC~\cite{smith2016practical} and \textsf{t$|$ket$\rangle$}~\cite{CQCdepth} compilers. The former uses CZ gates as basic two-qubit gates instead of CNOT gates. Since a CNOT gate can be formed by conjugating the target bit of a CNOT gate with Hadamards, the amount of CZ gates can be compared directly to the amount of CNOT gates. 

As architectures, we used a 9-qubit square grid, the 20-qubit IBM Q20 Tokyo, a 16-qubit square grid, the IBM QX5 and  the Rigetti 16Q Aspen architectures. Their respective gate counts and percentage of added gates (overhead) can be found in
Table~\ref{table:results}
\footnote{All circuits can be found in QASM format at: \\\href{https://github.com/Quantomatic/pyzx/tree/steiner\_decomp/circuits/steiner}{https://github.com/Quantomatic/pyzx/tree/steiner\_decomp/circuits/steiner}}. For the genetic algorithm, we used different parameters depending on the size of the architecture. For the 9-qubit architecture, we used a population of $30$ and $15$ iterations. For the 16-qubit architectures, we used a population of $50$ and $100$ iterations. And for the 20-qubit architecture, we used a population of $100$ and $100$ iterations. The constant crossover and mutation probability had a constant value of $0.8$ and $0.2$, respectively. The values population and iteration were simply found by trial and error, and could probably be tuned further. A larger population and more iterations improves the chances of finding a better permutation, but increases the time that the algorithm needs to run. Our running times range from a few seconds to about a minute on a mid-range laptop computer from 2017.

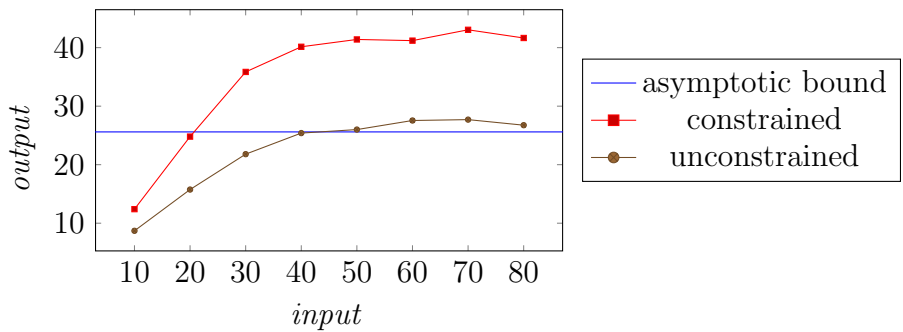
\begin{figure}
\centering
\begin{tikzpicture}[scale=2]
\begin{axis}[
width=14cm,
height=8cm,
x label style={at={(axis description cs:0.9,-0.1)},anchor=north,font=\it},
y label style={at={(axis description cs:0,1.2)},anchor=east,font=\it},
xlabel=input,
ylabel=output,
legend style={at={(2,1)},anchor=west}
]
\addplot [blue, line legend,
sharp plot,update limits=false,
] coordinates { (1,25.6) (90,25.6) };
\addplot+ [
sharp plot,
] coordinates {
(10,12.4)
(20,24.8)
(30,35.85)
(40,40.15)
(50,41.4)
(60,41.2)
(70,43.05)
(80,41.65)
};
\addplot+ [
sharp plot,
] coordinates {
(10,8.7)
(20,15.75)
(30,21.8)
(40,25.4)
(50,26.0)
(60,27.55)
(70,27.7)
(80,26.75)
};

\legend{asymptotic bound, constrained, unconstrained}
\end{axis}
\end{tikzpicture}
\caption{\label{fig:stabilise} A plot of input vs. output CNOT gates when mapping on to a 9-qubit square grid, when computing and re-synthesising the circuit, both in the unconstrained case using the method described in~\cite{PatelCNOT} and respecting nearest-neighbour constraints using our method.}
\end{figure}

Note that the overheads of our mapping process are sometimes negative. This is because the process we use computes the parity map associated to a CNOT circuit and re-synthesises the circuit using Gaussian elimination, as described in Sections~\ref{sec:parity} and~\ref{sec:algorithm}. In the unconstrained case, Patel \textit{et al.}~\cite{PatelCNOT} gave a heuristic for (re-)synthesising generic CNOT circuits on $n$ qubits with $O(n^2/\log(n))$ CNOTs. Indeed, using their algorithm for 9-qubit circuits, we see in Fig.~\ref{fig:stabilise} that the average CNOT count stabilises around the asymptotic bound of $9^2/\log_2(9) \approx 25.6$. Our approach also stabilises, but at a higher number of CNOTs ($\sim42$ for 9 qubits).

As a point of comparison, a na\"ive mapping procedure, such as the baseline procedure described in~\cite{nash2019routing}, introduces an overhead of $4(d-1)$ CNOT gates per gate introduced by the Patel \textit{et al.} synthesis algorithm, where $d$ is the length of the shortest path between two qubits. Since the average Manhattan distance on a 9-qubit square is $d = 2$, we expect the na\"ive method to stabilise at $25.6 \cdot 4 \cdot (2 - 1) \approx 102$ CNOT gates on the 9-qubit square. 


\section{Extensions}\label{sec:extensions}

\subsection{Mapping to general graphs}\label{sec:gen-graphs}

We can extend to graphs that do not contain a Hamiltonian path by using a recursive version of the algorithm described in Section~\ref{sec:algorithm}. For this, we will define parametrised versions, $\steinerdown(k, V)$ and $\steinerup(k, V, W)$, of the procedures from Section~\ref{sec:algorithm}. In both cases, the procedures now take a parameter $k$ which gives the current vertex/column on which to perform the downward or upward Gaussian elimination, respectively. They are also given a set $V \subseteq V_G$ over which all operations are restricted. Hence, rather than progressively deleting vertices from $G$, we start with $V := V_G$ and progressively remove vertices from $V$. The second procedure, \steinerup, also has a parameter $W \subseteq V$ where the Steiner tree and row operations are allowed to be non-descending. That is, all edges in the rooted Steiner tree $T$ must be descending unless both vertices adjacent to that edge are in $W$. Similarly, we require that row operations should be descending unless the rows corresponding to both vertices are in $W$.

Let $G$ be an undirected graph and $R$ a spanning tree.
Fix some leaf of $R$ and number the vertices consecutively by depth-first traversal (DFT), ensuring that each vertex is labelled \textit{after} its children (i.e. post-ordering). We then choose vertex $0$ as the root of $R$. Such a numbering has the property that removing vertices in ascending order never results in a disconnected graph. Here are some examples:
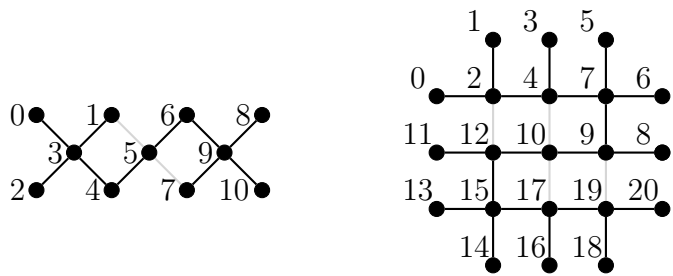
\[
\begin{tikzpicture}
	\begin{pgfonlayer}{nodelayer}
		\node [style=black vertex] (22) at (-2, 0) {};
		\node [style=black vertex] (23) at (-1, 1) {};
		\node [style=black vertex] (24) at (0, 0) {};
		\node [style=black vertex] (25) at (-1, -1) {};
		\node [style=black vertex] (26) at (1, 1) {};
		\node [style=black vertex] (27) at (2, 0) {};
		\node [style=black vertex] (28) at (3, 1) {};
		\node [style=black vertex] (29) at (3, -1) {};
		\node [style=black vertex] (30) at (1, -1) {};
		\node [style=black vertex] (31) at (-3, 1) {};
		\node [style=black vertex] (32) at (-3, -1) {};
		\node [style=none] (33) at (-3.5, 1) {$0$};
		\node [style=none] (34) at (-3.5, -1) {$2$};
		\node [style=none] (35) at (-1.5, 1) {$1$};
		\node [style=none] (36) at (-2.5, 0) {$3$};
		\node [style=none] (37) at (-1.5, -1) {$4$};
		\node [style=none] (38) at (-0.5, 0) {$5$};
		\node [style=none] (39) at (0.5, 1) {$6$};
		\node [style=none] (40) at (1.5, 0) {$9$};
		\node [style=none] (41) at (2.5, 1) {$8$};
		\node [style=none] (42) at (0.5, -1) {$7$};
		\node [style=none] (43) at (2.25, -1) {$10$};
	\end{pgfonlayer}
	\begin{pgfonlayer}{edgelayer}
		\draw [style=black edge] (31) to (22);
		\draw [style=black edge] (22) to (25);
		\draw [style=black edge] (25) to (24);
		\draw [style=black edge] (24) to (26);
		\draw [style=black edge] (26) to (27);
		\draw [style=black edge] (27) to (29);
		\draw [style=black edge] (32) to (22);
		\draw [style=black edge] (22) to (23);
		\draw [style=black edge] (30) to (27);
		\draw [style=black edge] (27) to (28);
		\draw [style=gray edge] (23) to (24);
		\draw [style=gray edge] (24) to (30);
	\end{pgfonlayer}
\end{tikzpicture}
\qquad\qquad
\begin{tikzpicture}
	\begin{pgfonlayer}{nodelayer}
		\node [style=black vertex] (0) at (0, 1.5) {};
		\node [style=black vertex] (1) at (1.5, 1.5) {};
		\node [style=black vertex] (2) at (-1.5, 1.5) {};
		\node [style=black vertex] (3) at (-1.5, 0) {};
		\node [style=black vertex] (4) at (-1.5, -1.5) {};
		\node [style=black vertex] (5) at (0, -1.5) {};
		\node [style=black vertex] (6) at (1.5, -1.5) {};
		\node [style=black vertex] (7) at (1.5, 0) {};
		\node [style=black vertex] (8) at (0, 0) {};
		\node [style=black vertex] (9) at (-3, 1.5) {};
		\node [style=black vertex] (10) at (-1.5, 3) {};
		\node [style=black vertex] (11) at (0, 3) {};
		\node [style=black vertex] (12) at (1.5, 3) {};
		\node [style=black vertex] (13) at (3, 1.5) {};
		\node [style=black vertex] (14) at (3, 0) {};
		\node [style=black vertex] (15) at (3, -1.5) {};
		\node [style=black vertex] (16) at (1.5, -3) {};
		\node [style=black vertex] (17) at (0, -3) {};
		\node [style=black vertex] (18) at (-1.5, -3) {};
		\node [style=black vertex] (19) at (-3, -1.5) {};
		\node [style=black vertex] (20) at (-3, 0) {};
		\node [style=black vertex] (21) at (3, 0) {};
		\node [style=none] (22) at (-3.5, 2) {$0$};
		\node [style=none] (23) at (-2, 3.5) {$1$};
		\node [style=none] (24) at (-2, 2) {$2$};
		\node [style=none] (25) at (-0.5, 3.5) {$3$};
		\node [style=none] (26) at (-0.5, 2) {$4$};
		\node [style=none] (27) at (1, 3.5) {$5$};
		\node [style=none] (28) at (1, 2) {$7$};
		\node [style=none] (29) at (2.5, 2) {$6$};
		\node [style=none] (30) at (-3.5, 0.5) {$11$};
		\node [style=none] (31) at (-2, 0.5) {$12$};
		\node [style=none] (32) at (-0.5, 0.5) {$10$};
		\node [style=none] (33) at (1, 0.5) {$9$};
		\node [style=none] (34) at (2.5, 0.5) {$8$};
		\node [style=none] (35) at (-3.5, -1) {$13$};
		\node [style=none] (36) at (-2, -1) {$15$};
		\node [style=none] (37) at (-0.5, -1) {$17$};
		\node [style=none] (38) at (1, -1) {$19$};
		\node [style=none] (39) at (2.5, -1) {$20$};
		\node [style=none] (40) at (-2, -2.5) {$14$};
		\node [style=none] (41) at (-0.5, -2.5) {$16$};
		\node [style=none] (42) at (1, -2.5) {$18$};
	\end{pgfonlayer}
	\begin{pgfonlayer}{edgelayer}
		\draw [style=black edge] (2) to (0);
		\draw [style=black edge] (0) to (1);
		\draw [style=black edge] (1) to (7);
		\draw [style=black edge] (7) to (8);
		\draw [style=black edge] (8) to (3);
		\draw [style=black edge] (3) to (4);
		\draw [style=black edge] (4) to (5);
		\draw [style=black edge] (5) to (6);
		\draw [style=black edge] (2) to (10);
		\draw [style=black edge] (2) to (9);
		\draw [style=black edge] (3) to (20);
		\draw [style=black edge] (4) to (19);
		\draw [style=black edge] (4) to (18);
		\draw [style=black edge] (5) to (17);
		\draw [style=black edge] (6) to (16);
		\draw [style=black edge] (6) to (15);
		\draw [style=black edge] (7) to (21);
		\draw [style=black edge] (1) to (13);
		\draw [style=black edge] (1) to (12);
		\draw [style=black edge] (0) to (11);
		\draw [style=gray edge] (2) to (3);
		\draw [style=gray edge] (0) to (8);
		\draw [style=gray edge] (8) to (5);
		\draw [style=gray edge] (7) to (6);
	\end{pgfonlayer}
\end{tikzpicture}
\]
Note that, since we use a post-ordering, the starting point of the DFT will \textit{not} end up as the root of $R$. In the graphs above, the starting points for the DFT end up labelled $10$ and $20$, respectively, whereas the root of $R$ is always taken to be $0$.

For a parity map $P$ whose rows are labelled by the vertices of $G$, the recursive version of the previous procedure, called \steinerrec, is defined as follows:
\begin{enumerate}
  \item For each $k \in G$ (in ascending order), apply $\steinerdown(k, \{j \,|\, j \geq k \})$.
  \item If $R$ is empty, we are done. Otherwise pick the maximal vertex $k \in R$ and maximal leaf $k' \in R$.
  \item Let $W$ be the set of vertices in the shortest path from $k'$ to $k$ (inclusive). Apply $\steinerup(k', V_R, W)$.
  \item Apply \steinerrec on the subgraph of $G$ restricted to $W$.
  \item Remove $k'$ from $R$ and go to 2.
\end{enumerate}
Note that, when the spanning tree $R$ is actually a Hamiltonian path, it will always be the case that $k = k'$, hence the recursive part is trivial and it reduces to the algorithm described in Section~\ref{sec:algorithm}. If $k$ is not a leaf, note that, because of our choice of numbering, the set $W$ is maximal in $V_R$ with respect to $\leq$. This corresponds to a block matrix $B$ in the bottom-right corner of the region of $P$ which is not yet `finished'. That is, $P$ decomposes as:
\[
P = \left( \  \begin{tikzpicture}
	\begin{pgfonlayer}{nodelayer}
		\node [style=none] (0) at (-3.5, 2.5) {};
		\node [style=none] (1) at (-3.5, -2.5) {};
		\node [style=none] (2) at (4.5, 2.5) {};
		\node [style=none] (3) at (4.5, -2.5) {};
		\node [style=none] (5) at (-3.5, -1) {};
		\node [style=none] (6) at (2.5, 2.5) {};
		\node [style=none] (7) at (2.5, -1) {};
		\node [style=none] (8) at (0.5, 0.5) {};
		\node [style=none] (9) at (0.5, -1) {};
		\node [style=none] (10) at (2.5, 0.5) {};
		\node [style=none] (11) at (-0.5, 1.5) {$P'$};
		\node [style=none] (13) at (1.5, -0.25) {$B$};
		\node [style=none] (14) at (2.5, -2.5) {};
		\node [style=none] (15) at (4.5, -1) {};
		\node [style=none] (16) at (3.5, 0.75) {$0$};
		\node [style=none] (17) at (-0.5, -1.75) {$0$};
		\node [style=none] (18) at (3.5, -1.75) {$1$};
		\node [style=none] (19) at (-3.5, 0.5) {};
		\node [style=none] (20) at (-1.5, -0.25) {$0$};
	\end{pgfonlayer}
	\begin{pgfonlayer}{edgelayer}
		\draw [style=dashed edge] (0.center) to (2.center);
		\draw [style=dashed edge] (2.center) to (3.center);
		\draw [style=dashed edge] (1.center) to (3.center);
		\draw [style=dashed edge] (0.center) to (1.center);
		\draw [style=dashed edge] (6.center) to (7.center);
		\draw [style=dashed edge] (5.center) to (7.center);
		\draw [style=dashed edge] (8.center) to (10.center);
		\draw [style=dashed edge] (8.center) to (9.center);
		\draw [style=dashed edge] (7.center) to (15.center);
		\draw [style=dashed edge] (7.center) to (14.center);
		\draw [style=dashed edge] (19.center) to (8.center);
	\end{pgfonlayer}
\end{tikzpicture} \  \right)
\]
Note there are zeroes to the left of $B$ because step 1 made $P$ upper-triangular. Step 3 makes everything above $P_{k',k'}$ into 0, but it might mess up the upper-triangular structure of $B$ itself, since we allow non-decreasing row operations between vertices in $W$. Hence, the recursive call in step 4 fixes $B$ up afterwards. At this point, the $k'$-th row and column only contain a single $1$ and since it is a leaf, $k'$ will not be needed for another Steiner tree. Hence it is `finished', so we can remove it from $R$. Note that, in successive iterations of this procedure, $B$ might not necessarily consist of consecutive rows/columns, but it will always contain all of the maximal `unfinished' rows/columns in $R$. Hence, the reasoning is identical.

Since each iteration of steps 2-5 removes a vertex from $R$ and the recursive call in step 4 is always restricted to a proper subgraph of the current graph, the procedure terminates when $R$ is empty and $P$ is in reduced echelon form.

\subsection{Mapping from general circuits}\label{sec:gen-circuits}

We can also extend the technique we proposed for CNOT circuits to more general families of circuits: namely CNOT+Rz circuits and even generic Clifford+Rz circuits. We first explain the simpler case, as it does not require going outside of the circuit model. By Rz gates, we mean a Z-phase rotation by a generic phase $\alpha$:
\[ 
\Rz{\alpha} := \left(
\begin{matrix}
  1 & 0 \\
  0 & e^{i\alpha} \\
\end{matrix}
\right)
\]
Certain special cases have common names in quantum circuit literature, e.g. $S := \Rz{\frac\pi2}$ and $T := \Rz{\frac\pi4}$. As explained in e.g.~\cite{amy2014polynomial}, circuits consisting of CNOT gates and arbitrary Z-phase rotations can be described efficiently in two parts: a \bitfield-linear map describing the action of the circuit on basis states and a set of pairs of the form $(\alpha, v)$ where $\alpha \in [0, 2\pi)$ is an angle and $v$ is a vector in \bitfield describing the \textit{parity} of input states on which that angle is applied.

First, note that it is straightforward to efficiently calculate the behaviour of CNOT+Rz circuits on computational basis states. One simply labels the input wires by variables $(x_0, \ldots, x_{n-1})$ and propagates these labels from left-to-right using the rule that $\Rz{\alpha}$ does not change the labels, whereas CNOT sends labels $(x,y)$ on its inputs to labels $(x, x\oplus y)$ on its outputs:
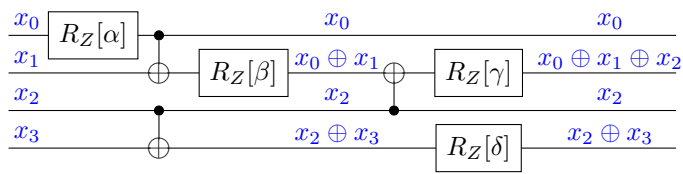
\ctikzfig{cnot-phase-circuit-marked}
We can then read off the output basis state from the final labels on the wires and the parities associated with each phase from the wire that the phase gate is on. For example, the circuit above acts as follows on computational basis states:
\begin{equation}\label{eq:phase-poly}
\ket{x_0, x_1, x_2, x_3} \mapsto e^{i [\alpha \cdot x_0 + \beta \cdot (x_0 \oplus x_1) + \gamma \cdot (x_0 \oplus x_1 \oplus x_2) + \delta (x_2 \oplus x_3)]} \ket{x_0, x_0 \oplus x_1 \oplus x_2, x_2, x_2 \oplus x_3}
\end{equation}
for all $x_i \in \{0,1\}$. The expression $\phi$ in $e^{i\phi}$ above dictates how the phase depends on the input basis state, and is sometimes referred to as a \textit{phase polynomial}. Clearly the relevant data for the overall unitary is the (\bitfield-linear) action on basis states as well as this phase polynomial.
We can then represent a parity of input variables as a bitstring, e.g. $x_0 \oplus x_2 \oplus x_3$ is the bitstring $(1, 0, 1, 1)$, indicating this parity depends on the first, third, and forth input variables. Hence data associated with the map \eqref{eq:phase-poly} is a parity matrix $P$ action on basis states and a set of angle/vector pairs $\mathcal P$ giving the terms of the phase polynomial:
\[
\left(
P = \left(
\begin{matrix}
  1 & 0 & 0 & 0 \\
  1 & 1 & 1 & 0 \\
  0 & 0 & 1 & 0 \\
  0 & 0 & 1 & 1 \\
\end{matrix}
\right),
\ \ 
\mathcal P = \left\{
\alpha \cdot
\left(
\begin{matrix}
  1 \\
  0 \\
  0 \\
  0
\end{matrix}
\right), \ 
\beta \cdot
\left(
\begin{matrix}
  1 \\
  1 \\
  0 \\
  0
\end{matrix}
\right), \ 
\gamma \cdot
\left(
\begin{matrix}
  1 \\
  1 \\
  1 \\
  0
\end{matrix}
\right), \ 
\delta \cdot
\left(
\begin{matrix}
  0 \\
  0 \\
  1 \\
  1
\end{matrix}
\right)
\right\}
\ \ 
\right)
\]
Much like the CNOT case, there is an evident circuit extraction procedure based on Gaussian elimination for CNOT+Rz maps. One can use CNOT gates to perform primitive row operations on sets of parity vectors to reduce them to unit vectors, which then correspond to applying \Rz{\alpha} gates on single qubits. If this Gaussian elimination is performed using the \steinergauss procedure, these primitive row operations, and hence the synthesised circuit, will respect nearest-neighbour constraints, just as in the CNOT case.



A very similar procedure based on Gaussian elimination is described in a recent article by one of the authors~\cite{cliff-simp} as a means of extracting a quantum circuit from a simplified tensor-network-like representation called a \textit{ZX-diagram}. The extraction phase proceeds from right-to-left on a directed acyclic graph, and exploits the fact that post-composing CNOT gates is able to perform primitive row operations on the adjacency matrix of the graph:
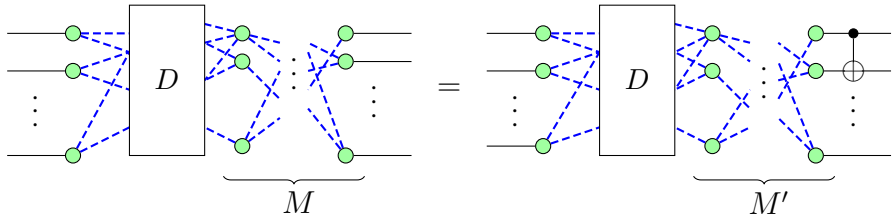
\ctikzfig{cnot-pivot}
This rule follows from a graph-theoretic transformation on ZX-diagrams called \textit{pivoting}. While the details of how this actually works are not relevant here, an important observation is that this procedure makes use of Gaussian elimination to produce CNOT gates, so substituting the \steinergauss algorithm immediately gives a circuit extraction procedure that respects nearest-neighbour constraints of the architecture. Since the technique described in \cite{cliff-simp}, and the corresponding PyZX circuit optimisation tool~\cite{pyzx}, take an arbitrary quantum circuit as input, this will give a general-purpose, optimising routine for circuit mapping. We leave a detailed exposition (and implementation) of this technique for future work.

\section{Conclusions and Future Work}

We have demonstrated a CNOT circuit mapping procedure that significantly out-performs existing compilers on memory architectures whose graphs contain a Hamiltonian path. We have also outlined extensions of this technique to arbitrary graphs and to arbitrary circuits. In addition to implementing the extensions outlined in the previous section, an interesting direction for future work is to focus not only on decreasing two-qubit gate count, but also in decreasing gate depth and/or overall fidelity loss due to gate errors. With the incredibly short coherence times characteristic of superconducting hardware, gate depth is likely to play an even more important role than gate count to practical realisation of quantum computations in the coming years~\cite{CQCdepth}. It also comes with a unique set of challenges, as parallel gates can interfere with each other if they are mapped to neighbouring locations on some architectures~\cite{googleupdate}. Another notable feature of current hardware is not all qubits are created equal: performing 2-qubit gates between certain pairs of qubits can be done with much higher fidelities than others, depending on implementation details or even random variance in the manufacturing processes of superconducting chips~\cite{regittiparams}. While the optimisation techniques used in this paper are very simple, a topic of future work is to apply much more powerful machine learning and/or constraint satisfaction techniques in order to take these factors into account.

\paragraph{Acknowledgements.} We gratefully acknowledge support from the Unitary Fund (\url{http://unitary.fund}) for this work. We would also like to thank Will Zeng, Ross Duncan, and John van de Wetering for fruitful discussions about circuit mapping for NISQ as well as the authors of \cite{nash2019routing} for clarifying some points about their approach.

\newpage


\section*{References}

\bibliographystyle{unsrt}
\bibliography{quantum}

\end{document}